
\input amstex
\magnification 1200
\documentstyle{amsppt}
\NoBlackBoxes
\NoRunningHeads
\def\Fun{\text{Fun}}
\def\C{\Bbb C}

\def\N{\Bbb N}
\def\Z{\Bbb Z}

\def\<{\langle}
\def\>{\rangle}

\def\Rep{\text{Rep}}
\def\RR{(\text{Rep}\Gamma)^{\Sigma}}
\def\G{\Gamma}
\def\S{\Sigma}
\def\Re{\text{Res}_{\Gamma}^G(V)}
\topmatter
\title Representations of tensor categories and Dynkin diagrams
\endtitle
\endtopmatter
\document

\centerline {\bf Pavel Etingof\footnote
{   Department of Mathematics,
   Harvard University,
   Cambridge, MA 02138, USA,\linebreak
    e-mail: etingof\@math.harvard.edu		}
 and Michael Khovanov\footnote{   Department of Mathematics,
   Yale University,
   New Haven, CT 06520, USA,\linebreak
   e-mail: michaelk\@math.yale.edu		}
}
\vskip .25in
\centerline{August 12, 1994}
\centerline{hep-th 9408078}
\vskip .15in

In this note we illustrate by a few examples the general
principle: interesting
algebras and representations defined over $\Z_+$
come from category theory, and are best understood
when their categorical origination has been discovered.
We show that indecomposable $\Z_+$-representations of the character
ring of $SU(2)$ satisfying certain conditions correspond to affine and
infinite Dynkin
diagrams with loops. We also show that irreducible
$\Z_+$-representations of the Verlinde algebra
(the character ring of the quantum group
$SU(2)_q$, where $q$ is a root of unity),
satisfying similar conditions correspond to usual
(non-affine) Dynkin diagrams with loops.
\vskip .03in

{\bf 1. $\Z_+$-representations.}

Let $\Z_+$ be the set of nonnegative integers.

\proclaim{Definition 1.1}
(i) \it  A $\Z_+$-basis \rm of an algebra $A$ over $\C$ is a basis $B=\lbrace
b_i\rbrace$ such that
$b_ib_j=\sum_k c_{ij}^k b_k,\ c_{ij}^k\in \Z_+.$

(ii) \it A $\Z_+$-algebra \rm is a $\C$-algebra with unit having a fixed
$\Z_+$-basis containing the unit.

(iii) \it A $\Z_+$-representation \rm of a
$\Z_+$-algebra $A$
is a $\C$-representation $M$ of $A$ with a basis $\lbrace m_j\rbrace$
such that
$b_im_j=\sum D_{ij}^k m_k,\ D_{ij}^k\in \Z_+.$

(iv) Two $\Z_+$-representations $M_1,M_2$ of $A$
with bases $\lbrace m_i^1\rbrace_{i\in I}$, $\lbrace m_j^2\rbrace_{j\in
J}$  are \it equivalent \rm iff there exists a bijection
$\phi:I\to J$ such that the induced linear mapping
$\tilde{\phi}$ of $\C$-vector spaces $M_1,M_2$ defined by
$\tilde{\phi}(m^1_i)=m^2_{\phi(i)}$
is an isomorphism of $A$-modules.

(v) \it The direct sum \rm of two
$\Z_+$-representations $M_1,M_2$ of $A$ is the representation $M_1\oplus
M_2$ of $A$ with the $\Z_+$-basis being the union of the $\Z_+$-bases of
$M_1$ and $M_2$.

(vi) A $\Z_+$-representation $M$ of $A$ is \it
indecomposable \rm if it is not equivalent to a direct sum of two nontrivial
$\Z_+$-representations.
\endproclaim

Examples of $\Z_+$-algebras and their $\Z_+$-representations
can be obtained by the following well known categorical construction.
Let $\Cal A,\Cal M$ be strict additive categories in which every
object can be uniquely, up to order, written as a
direct sum of indecomposable objects. Assume that $\Cal A$
is a tensor category (i.e. there is a biadditive functor
$\otimes: \Cal A\times\Cal A\to\Cal A$), and $\Cal M$ is a module
category over $\Cal A$ (i.e. there is a biadditive functor
$\otimes:\Cal A\times\Cal M\to\Cal M$). We assume that both tensor
product functors are associative, i.e. satisfy the pentagon axiom
\cite{Ma}. Then ones says that the category $\Cal M$ is a module
category over the ring category $\Cal A$.
Let $A,M$ be the Grothendieck groups of $\Cal A$, $\Cal M$
tensored with $\C$. Then $A$ is naturally a $\Z_+$-algebra, and
$M$ is a $\Z_+$-representation of $A$. The $\Z_+$-bases in $A,M$
are formed by indecomposable objects in $\Cal A$, $\Cal M$.

1.2. Let us consider an example.
Take a group $G$ and consider the $\Z_+$-algebra
 $A=\C [G]$ (the group algebra of $G$)
 with the $\Z_+$-basis $B=\lbrace g|g\in G\rbrace.$

If $H$ is a subgroup of finite index
in $G$, $X=G/H$, then the vector space $\C [X]$ with the
basis $\lbrace x|x\in X\rbrace$ is a $\Z_+$-representation of $\C [G]$.

\proclaim{Proposition 1.1}

(i) Any finite dimensional indecomposable
$\Z_+$-representation of $\C [G]$ is of the form $\C [G/H]$.

(ii) For two finite homogeneous $G$-spaces $X_1$ and $X_2$ the corresponding
$\Z_+$-representations $\C[X_1]$, $\C[X_2]$
are equivalent iff the action of $G$ on
$X_1$ is conjugate to the action of $G$ on $X_2$.
\endproclaim

\it Proof. \rm  If $n\times n$ matrices $T, T^{-1}$ have nonnegative
integer entries, it is easy to see that $T$ is a permutation matrix.
This implies (i). (ii) is obvious.
$\square$

Proposition 1.1 shows that equivalence classes of finite-dimensional
 $\Z_+$-representations
of $\C[G]$ are labeled by finite $G$-spaces up to congugacy.

All $\Z_+$-representations of $\C[G]$
can be constructed categorically. Let $\Fun G$
denote the algebra of complex functions on $G$ which vanish
outside of a finite set of elements. Let $\Fun X$ be the algebra of
complex functions on a $G$-set $X$. Then we have a homomorphism
$\Fun X\to \Fun G\otimes \Fun X$ associated to the action $G\times
X\to X$. This homomorphism defines a bifunctor $\otimes: \Rep\Fun G\times
\Rep\Fun X\to \Rep\Fun X$, where $\Rep A$ denotes the category
of finite dimensional representations of
an algebra $A$. At the level of Grothendieck groups
this functor yields the $\Z_+$-representation of $\C[G]$ in $\C[X]$.

Since $\C[G]$ is a Hopf algebra, the category of its
$\Z_+$-representations is a tensor category. It follows from
Proposition 1.1 that the Grothendieck ring of this category
coincides with the Burnside ring of $G$ (see \cite{CR}).
\vskip .03in

{\bf 2. $\Z_+$-representations of the character ring of
$SU(2)$.}

2.1. Let $G$ be a compact Lie group (for example, a finite group).
Consider the character ring   $(\text{Fun}G)^{G}$ of $G$ with the
$\Z_+$-basis $\lbrace \chi_V\rbrace$ consisting
of the characters of finite dimensional
irreducible representations of $G$. Thus, $(\text{Fun}G)^G$
is a $\Z_+$-algebra.
Examples of $\Z_+$-representations of $(\text{Fun}G)^G$ are obtained from
finite subgroups of $G$.

Let
$\Gamma\subset G$ be a finite subroup of $G$,
and let $\Rep G$, $\Rep \Gamma$ be the categories
 of finite dimensional representations of $G$, $\Gamma$.
We have a bifunctor $\otimes: \Rep G\times \Rep \Gamma\to \Rep\Gamma$:
$(V, U)\to \text{Res}_{\Gamma}^G(V)\otimes U.$
($\text{Res}$ denotes restriction).

Let $\Sigma$ be a subgroup of $\text{Aut}(\Rep \Gamma)$ such that
for any $\sigma\in \Sigma, V\in \Rep G , U\in \Rep \Gamma$,
$\Re\otimes \sigma(U)\cong \sigma(\Re\otimes U).$
Denote by $(\Rep\Gamma)^{\Sigma}$ the full subcategory  of $\Rep \Gamma$
whose objects are $\Sigma$-stable objects of $\Rep\Gamma$, i.e., such
objects $U$ that $\sigma(U)\cong U$.
$(\Rep\Gamma)^{\Sigma}$ is an additive category whose indecomposable
objects are orbital sums under the action of $\Sigma$
on the set of irreducible objects in
$\Rep\Gamma$. We have a unique decomposition in $\RR$ into the direct
sum of indecomposable objects, and the bifunctor $\otimes$ maps
$\Rep G\times (\Rep\Gamma)^{\Sigma}$ to $(\Rep\Gamma)^{\Sigma}$.
In particular, this bifunctor defines a $\Z_+$-representation of
$(\Fun G)^G$ in the Grothendieck group $M$ of $(\Rep\Gamma)^{\Sigma}$
given by $\chi_V[U]=[\Re \otimes U]$, where $[U]$ denotes the class
of the object $U$ in the Grothendieck group.
\vskip .03in

2.2. For example,  let $G=SU(2)$. Then $(\text{Fun}G)^{G}$ is spanned by
$\chi_1=1,\chi_2,\chi_3,...$ with
$\chi_i\chi_j=\sum_{k=|i-j|+1,k-|i-j|\text{ odd}}^{i+j-1}\chi_k$
(Clebsch-Gordan rule).

Let $\G\subset SU(2)$ be a finite subgroup, $V_1,...,V_r$ -- all irreducible
representations of $\G$. Define the matrix $\{\alpha_{ij}\}$ by
$\chi_2 \chi_{V_i}=\sum_j\alpha_{ij}\chi_{V_j}$, and set
$a_{ij}=2\delta_{ij}-\alpha_{ij}$. It was observed by McKay that
$\lbrace a_{ij}\rbrace$ is a simply laced affine Cartan matrix (see \cite{Mc}).

More precisely, this matrix is of the type:

$\tilde{A_n}$, $n\ge 1$ if $\G$ is a cyclic group
$\Z /(n+1)\Z \subset SU(2)$,

 $\tilde{D_n}$, $n\ge 4$ if $\G$ is a dihedral group $Q_{4(n-2)}$,

$\tilde{E_6}$ if $\G =p^*(A_4)$, where $A_4\subset SO(3)$ is the group of
symmetries of the regular tetrahedron preserving the orientation,
and $p$ is the two-fold covering $SU(2)\to SO(3)$,

$\tilde{E_7}$ if $\G =p^*(S_4)$, where $S_4\subset SO(3)$ is the group of
symmetries of the cube preserving the orientation,

$\tilde{E_8}$ if $\G=p^*(A_5)$, where $A_5\subset SO(3)$ is the group of
symmetries of the regular icosahedron preserving the orientation.

An affine Cartan matrix can be encoded by an affine Dynkin diagram,
which is a graph defined by the condition that $\alpha_{ij}$ is its incidence
matrix. The group $\S$ can thus be interpreted
as a group of automorphisms of the affine Dynkin diagram.
Thus we can assign a representation of $(\text{FunG})^{G}$ to any pair
$(\G ,\S )$, where $\G $ is simply laced diagram of an affine type, and
$\S $ is a group of automorphisms of $\G $.
\vskip .03in

2.3. Let us describe the representations corresponding to nontrivial groups
$\Sigma$. This description is standard.
The representations are depicted by non-simply laced affine Dynkin diagrams,
possibly with loops, obtained as quotients of simply laced diagrams
by a group of automorphisms. Such diagrams are listed, for example,
 in \cite{HPR}, where we
borrow the notation for these diagrams (except that
we denote by $\tilde A_1$ the diagram
denoted by $\tilde A_{12}$ in \cite{HPR}, and by $\tilde A_{12}$
the diagram denoted there by $\tilde A_{11}$).
Note that the number of nodes
of the diagram equals to the value of the first subscript plus 1.

Quotients of $\tilde A_n$. If $\Sigma$ is generated by the reflection
of the diagram which preserves a vertex, then the quotient
diagram is $\widetilde{BL}_m$, $m\ge 1$, if $n=2m$, and $\tilde B_m$,
$m\ge 2$, if $n=2m-1$. If $n=2m+1$, and $\Sigma$ is generated by
the reflection which has no
fixed vertices, then the quotient is $\tilde L_m$, $m\ge 0$.

Quotients of $\tilde D_n$. If $n=2m$, $m\ge 3$, and $\Sigma$ is
generated by a reflection
which preserves the middle vertex, and no other vertices, then the
quotient is $\widetilde{BD}_m$. If $n=2m+1$, and
$\Sigma$ is generated by a reflection preserving no vertices,
then the quotient is $\widetilde{DL}_m$, $m\ge 2$. If $\Sigma$ is
generated by a reflection preserving all vertices but the four
endpoints, then the quotient is $\tilde C_{n-2}$. If $\Sigma$ is
generated by a reflection that preserves all vertices but two,
then the quotient is $\widetilde{CD}_{n-1}$. If $\Sigma$
has four elements and acts transitively on the endpoints,
then the quotient is $\widetilde{BC}_m$ if $n=2m+2$ ($m\ge 2$),
$\widetilde{CL}_m$ if $n=2m+3$, $m\ge 1$, and $\tilde{A}_{12}$ if
$n=4$. If $n=4$ and $\Sigma$ is generated by an automorphism of order
3, then the quotient is $\tilde G_{21}$.

Quotients of $\tilde E_6$. If $\Sigma$ is of order 2, the quotient is
$\tilde F_{41}$. If $\Sigma$ is of order 3, then the quotient is
$\tilde G_{22}$.

Quotients of $\tilde E_7$. $\Sigma$ has to be of order 2, and the
quotient is $\tilde{F}_{42}$.

Quotients of $\tilde E_8$. None.

This list exhausts all affine diagrams with loops from \cite{HPR}.
\vskip .03in

2.4. The same method applies to construction of infinite-dimensional
irreducible $\Z_+$-representations of $(\Fun SU(2))^{SU(2)}$.
For this purpose consider an infinite
closed subgroup $\Gamma\subset SU(2)$, and construct the category
$(\Rep\Gamma)^{\Sigma}$ as in Section 2.2. Passage to the Grothendieck
groups will produce an infinite-dimensional irreducible
 $\Z_+$-representation, in which the action of $\chi_2$ is given by
the matrix $2I-A$, where $A$ is an infinite Cartan matrix, and
$I$ is the identity matrix.
Such matrices are depicted by infinite Dynkin diagrams with
loops, which are listed in \cite{HPR}. Again, we preserve the
 notation of \cite{HPR} for these diagrams.

Let us describe the possible cases.

If $\Gamma=SU(2)$ itself, then $\Sigma$ has to be trivial,
and the obtained diagram is $A_{\infty}$
(a semiinfinite chain). It corresponds to the regular
representation of the character ring.

If $\Gamma=U(1)$ and $\Sigma$ is trivial, then the obtained
diagram is $A_{\infty}^{\infty}$ (a chain infinite in both
directions). If $\Sigma$ is generated by a nontrivial reflection of the chain
then the obtained diagram is $B_{\infty}$ if a vertex is preserved by
the reflection, and $L_{\infty}$ if no vertex is preserved.

If $\Gamma=U(1)\cup \biggl(\matrix 0&1\\ -1&0\endmatrix\biggr)U(1)$,
and $\Sigma$ is trivial, then the obtained diagram is $D_{\infty}$.
If $\Sigma$ is generated by a nontrivial reflection, then the obtained
diagram is $C_{\infty}$.

This exhausts all infinite Dynkin diagrams with loops listed in \cite{HPR}.
\vskip .03in

2.5. In the above examples,
the Dynkin diagram encodes the matrix of $\chi_2$. The rest of
matrices $(\chi_3,\chi_4,...)$ are obtained by the formula
$$\chi_n=P_n(\chi_2),\tag 2.1$$
where $P_n$ is \it the ultraspherical polynomial \rm defined by
$$P_n(2\text{cos}x)=\frac{\text{sin}nx}{\text{sin}x}.\tag 2.2$$

They can be computed recursively by
$$
P_1(x)=1,\ P_2(x)=x,\ P_{n+2}(x)=xP_{n+1}(x)-P_{n}(x), n\ge 1.\tag 2.3
$$

So, from the above we get

\proclaim{Proposition 2.2} If $X=2I-A$, where
$A$ is an affine or infinite Cartan matrix from the list of
\cite{HPR}, then for any $n\ge 1$ $P_n(X)$ has nonnegative entries.
\endproclaim

{\bf Remark.} This is not obvious apriori since $P_n$ has both positive and
negative coefficients.
\vskip .03in

{\bf 3. Representations of the Verlinde algebra.}

3.1. Fix a nonnegative integer $k$.

\proclaim{Definition 3.1} The Verlinde algebra $V_k$ is the
commutative $\Z_+$-algebra with the basis $\{\chi_i\}$, $1\le i\le
k+1$, and relations
$$
\chi_i\chi_j=\sum_{l=|i-j|+1,l-|i-j|\text{
odd}}^{\text{min}(i+j-1,2k+3-i-j)}
\chi_l.\tag 3.1
$$
(the truncated Clebsch-Gordan rule).
\endproclaim

The Verlinde algebra is the Grothendieck ring of a tensor category,
namely, the fusion category for the $SU(2)$-Wess-Zumino-Witten
conformal field theory, which explains its $\Z_+$-structure.
This was the way this algebra appeared
in the physical work of E.Verlinde \cite{V}.
By now, several mathematical definitions of the fusion category
are available, out of which we would like to mention two. One of them
says that it is the category of integrable, highest weight
modules over the affine Lie algebra $\widehat{\frak sl}_2$ at level
$k$, with a special tensor product, called fusion,
defined by means of the operator product expansion
 (see \cite{MS}).
The other says that it is the tensor category of finite-dimensional
representations of the quantum group $U_q({\frak sl}_2)$ at
$q=e^{\frac{\pi i}{k+2}}$ modulo representations
of zero quantum dimension (see \cite{RT}). These two definitions are
equivalent.
\vskip .03in

3.2. $V_k$ has the following (known) characterization.

\proclaim{Proposition 3.1}
(i) $P_{k+2}(\chi_2)=0$ in $V_k$, where $P_n$ is defined by (2.3).
Also, $\chi_j=P_j(\chi_2)$, $1\le j\le k+1$.

(ii) The assignment $x\to \chi_2$ defines an isomorphism of algebras
$\C[x]/P_{k+2}(x)\C[x]\to V_k$.
\endproclaim

\demo{Proof} (i) Consider the matrix of multiplication by $\chi_2$
in $V_k$ in the basis of $\chi_i$. It is easy to see that it
is equal to the incidence matrix of the Dynkin diagram $A_{k+1}$.
The eigenvalues of this matrix are well known and equal $2\text{cos}\frac{\pi
m}{k+2}$, $m=1,...,k+1$ (this is due to Lagrange). Therefore,
according to (2.2), the eigenvalues of $P_{k+2}(\chi_2)$ are all zero,
and since this matrix is symmetric, it equals to $0$.
The relation $\chi_j=P_j(\chi_2)$ follows from
$\chi_2\chi_j=\chi_{j+1}+\chi_{j-1}$, $2\le j\le k$, and (2.3).

(ii) It follows from (3.1) that $V_k$ is generated by $\chi_2$.
Therefore, the map in question is an epimorphism. It is also a monomorphism
because the dimensions are the same.
$\square$
\enddemo

3.3. Now let $A$ be the Cartan matrix corresponding to a \it finite \rm
Dynkin diagram, possibly with a loop. These include
the usual Dynkin diagrams of finite dimensional simple Lie algebras,
i.e. $A_n$, $B_n$, $C_n$, $D_n$, $E_6$, $E_7$, $E_8$, $F_4$, $G_2$,
and an additional series $L_n$, $n\ge 1$, of diagrams with a loop (see
\cite{HPR}; the Cartan matrix of $L_n$ differs from that for $A_n$
only by
the value of the entry $a_{11}$,
which is $1$ in the case of $L_n$, instead of $2$).
$L_n$ is obtained from $A_{2n}$ as a quotient by a reflection.

Let $h$ be the Coxeter number of $A$ (the Coxeter number of $L_n$ is
$2n+1$). Let $X=2I-A$.

\proclaim{Proposition 3.2} $P_h(X)=0$.
\endproclaim

\demo{Proof} This follows from the result of Coxeter \cite{Co}:
the eigenvalues of $X$ have the form $2\text{cos}\frac{\pi m_j}{h}$,
where $m_j$ are the exponents of $A$.
This statement, together with (2.2), implies
the Proposition. $\square$
\enddemo

{\bf Remark.} The exponents of $L_n$
are defined by the condition that the matrix $X_{L_n}$
is the restriction of the matrix $X_{A_{2n}}$ to an
invariant subspace, and hence the set of exponents of $L_n$ is a subset
of that for $A_{2n}$; they turn out to be equal to $1,3,...,2n-1$.

\proclaim{Corollary 3.3} Let $k+2=h$. Then the assignment $\chi_2\to
X$ defines a matrix representation of $V_k$.
\endproclaim

Denote this representation by $R_A$.

The following theorem is the main result of this section.

\proclaim{Theorem 3.4} The representation $R_A$ is a $\Z_+$-representation
in the basis of unit vectors.
\endproclaim

\demo{Proof} We need to show that the elements $\chi_i$ are realized
by matrices of nonnegative integers in $R_A$, i.e. that the matrices
$P_n(X)$ have nonnegative integer entries for $n=1,2,...,k+1$.
Integrality is obvious, we only need to prove nonnegativity.
Unfortunately, we do not know a uniform proof, so we do
it case by case.

Type $A_n$. In this case $R_A$ is the regular representation
of $V_k$, and nonnegativity follows from (3.1): it expresses
the fact that $V_k$ is a Grothendieck ring of a category.

Types $B_n$ and $L_n$. These cases follow from the cases $A_{2n-1}$,
$A_{2n}$, since $B_n$, $L_n$ are quotients of $A_{2n-1}$, $A_{2n}$ by
an automorphism of order 2.

Type $C_n$. This case follows from the case $B_n$, since
$X_{C_n}=\Lambda^{-1} X_{B_n}\Lambda$, where $\Lambda$ is a diagonal
matrix with nonnegative entries (all but one diagonal entries of $\Lambda$
are equal to $1$, and the remaining entry equals $2$).

Type $D_n$. Here we use the fact that the quotient of $D_n$ by the
nontrivial automorphism of order 2 is $C_{n-1}$. Let $1,2$ be the
labels of the vertices permuted by this automorphism (all the other
vertices are fixed). Then the first two rows and the first two columns
of $X_{D_n}$ are the same. The same is true for the matrix
$P_m(X_{D_n})$, i.e.
$$
P_m(X_{D_n})=\biggl( \matrix a&a&\bold v^t\\ a&a&\bold v^t\\ \bold
v&\bold v&Y\endmatrix\biggr),\tag 3.2
$$
$a\in\Z$, $\bold v\in\Z^{n-2}$, $Y\in \text{Mat}_{n-2}(\Z)$.
Since the restriction of $X_{D_n}$ to the subspace of vectors whose
first two components are the same is $X_{C_{n-1}}$, we have
$$
P_m(X_{C_{n-1}})=\biggl( \matrix 2a&\bold v^t\\ 2\bold
v&Y\endmatrix\biggr).\tag 3.3
$$
Since nonnegativity has been proved for $C_{n-1}$, and the Coxeter
numbers of $D_n$, $C_{n-1}$ are the same ($2n-2$), we find that
$a,\bold v, Y$ have nonnegative integer entries. Therefore, so does
$P_m(X_{D_n})$, $m<2n-2$.

Type $G_2$. Follows from the case $D_4$, since $G_2$ is a quotient of
$D_4$ by an automorphism of order 3.

Types $E_6$, $E_7$, $E_8$. These cases have been checked by a computer.

Type $F_4$. Follows from the case $E_6$, since $F_4$ is a quotient
of $E_6$ by an automorphism of order 2.
$\square$\enddemo

\proclaim{Open problem} Construct a module category
over the Wess-Zumino-Witten fusion category which
would produce the $\Z_+$-representation $R_A$ at the level of
Grothendieck groups, and thus find a conceptual proof of Theorem 3.4.
\endproclaim

Clearly, it is enough to solve this problem for simply laced diagrams
of type $A,D,E$, since the rest can be obtained as quotients.

So far we only know how to construct such a category for
Dynkin diagrams of type $A_n$ and their quotients
$B_n$, $L_n$ (using the automorphism construction from Section 2.1).
The case $C_n$ can be obtained from $B_n$ as a subcategory.
\vskip .03in

{\bf 4. Classification of $\Z_+$-representations.}

4.1. $\Z_+$-representations $M$ of $(\Fun SU(2))^{SU(2)}$ constructed in
Chapter 2 have the following two properties (i), (ii).

(i) There exists a vector $\bold d$ over $\N$ with
$\bold d\chi_j=j\bold d$.

Indeed, looking at the construction of Section 2.2,
it is easy to see that the components of $\bold d$
are equal to the dimensions of indecomposable objects
in $\RR$, and the equation $\bold d\chi_j=j\bold d$ says that
$\text{dim}(V\otimes U)=\text{dim}V\text{dim}U.$

(ii) There exists an inner product $<,>$ on $M$ such that
$<m_i,m_j>=0, i\ne j$ and $\chi_i^*=\chi_i$ w.r.t. $<,>$.

This product
in Chapter 2 is given by $<U_1,U_2>=\text{dim}\text{Hom}_{\G}(U_1,U_2)$,
and the equation $\chi_i^*=\chi_i$ expresses the facts:
$V^*=V, V\in \Rep SU(2)$,
$\text{Hom}_{\G}(U_1, V\otimes U_2)=\text{Hom}_{\G}(V^*\otimes U_1,U_2)$.

\proclaim{Theorem 4.1}

(i) Any indecomposable finite-dimensional
$\Z_+$-representation of
$(\text{Fun}SU(2))^{SU(2)}$ with properties (i),(ii) is equivalent to
a representation defined by an affine Dynkin diagram (see Section 2.3).

(ii) Any indecomposable infinite-dimensional
$\Z_+$-representation of
$(\text{Fun}SU(2))^{SU(2)}$ with properties (i),(ii) is equivalent to
a representation defined by an infinite Dynkin diagram (see Section 2.3).
\endproclaim

\it Proof. \rm It is clear that every indecomposable representation
is countably dimensional.

Let $X$ be the matrix of $\chi_2$ and let $\bold d$ be
the vector with positive entries satisfying
$\bold dX=2\bold d$.
Consider the matrix $A=2I-X$.
Components $a_{ij}$ of $A$ are integers and satisfy

(a) $a_{ii}\le 2$,

(b) $a_{ij}\le 0$ if $i\ne j$,

(c) $a_{ij}=0$ if and only if $a_{ji}=0$ (because of property (ii)),

(d) $\bold d A=0$.

It was proved in \cite{HPR}
 that any such indecomposable integer
matrix $A$ is an affine Cartan matrix if it is of a finite size,
and an infinite Cartan matrix if it is of an infinite size.
(the finite size case was earlier done by Vinberg \cite{Vi}).
$\square$

\sl
Thus, we have shown that indecomposable $\Z_+$-representations of $(\Fun
SU(2))^{SU(2)}$ are in one-to-one correspondence with
affine and infinite Dynkin diagrams.
\rm
\vskip .03in

4.2. $\Z_+$-representations $M=R_A$ of the Verlinde algebra
$V_k$ constructed in Chapter 3 have the following two properties (i),(ii).

(i) There exists a vector $\bold d$ with positive real entries such that
$\bold d\chi_j=[j]_q\bold d$, where
$[j]_q=\frac{q^j-q^{-j}}{q-q^{-1}}$,
$q=e^{\frac{\pi i}{k+2}}$.

Indeed, by the result of Coxeter mentioned in the proof of
Proposition 3.2, the largest eigenvalue of the matrix $X$
is $[j]_q$, so $\bold d$ is just the left Perron-Frobenius
eigenvector of the matrix $X$. These eigenvectors for classical
Cartan matrices are listed in \cite{GHJ}, Table 1.4.8.
In the cases when we have a categorical construction for $R_A$
(i.e. cases $A_n$, $B_n$, $C_n$, $L_n$), the components of $\bold d$
have a representation-theoretic meaning -- they are the quantum
dimensions of the indecomposable objects
as representations of $U_q({\frak sl}_2)$, (see \cite{RT}), and
the equation $\bold d\chi_j=[j]_q\bold d$ says that
$\text{dim}_q(V\otimes U)=\text{dim}_qV\text{dim}_qU.$
($\text{dim}_q$ is the quantum dimension).

(ii) There exists an inner product $<,>$ on $M$ such that
$<m_i,m_j>=0, i\ne j$ and $\chi_i^*=\chi_i$ w.r.t. $<,>$.

The existence of $<,>$ follows from the symmetrizability of $A$,
i.e from the existence of a diagonal matrix $\Lambda$ such that $A^t=
\Lambda^{-1} A\Lambda$. In the cases $A_n,B_n,C_n,L_n$, the form $<,>$
 also has
a representation-theoretic meaning. Namely, it
is given by $<U_1,U_2>=\text{dim}\text{Hom}_{U_q({\frak sl}_2)}(U_1,U_2)$,
and the equation $\chi_i^*=\chi_i$ expresses the facts:
$V^*=V$ when $V\in \Rep U_q({\frak sl}_2)$ is irreducible, and
$\text{Hom}_{U_q({\frak sl}_2)}
(U_1, V\otimes U_2)=\text{Hom}_{U_q({\frak sl}_2)}(V^*\otimes U_1,U_2)$.

\proclaim{Theorem 4.2}
 Any indecomposable
$\Z_+$-representation of $V_k$ with properties (i),(ii) is equivalent to
a representation $R_A$ defined by a Dynkin diagram of finite type
with Coxeter number $h=k+2$ (see Section 3.3).
\endproclaim

\it Proof. \rm
Let $X$ be the matrix of $\chi_2$ and let $\bold d$ be
the vector with positive entries satisfying
$\bold dX=[2]_q\bold d$.
Consider the matrix $A=2I-X$.
Components $a_{ij}$ of $A$ are integer and satisfy

(a) $a_{ii}\le 2$,

(b) $a_{ij}\le 0$ if $i\ne j$,

(c) $a_{ij}=0$ if and only if $a_{ji}=0$ (because of property (ii)),

(d) $\bold d A>0$ componentwise (because
$2>[2]_q=2\text{cos}{\frac{\pi}{k+2}}$).

It was proved in \cite{HPR} that any such indecomposable integer
matrix $A$ is of a finite size and coincides with
a Cartan matrix for a Dynkin diagram of finite type.
The equation $P_h(X)=0$ implies that $h\le k+2$.
Also, it is clear that certain entries of $P_{h+1}(X)$ are negative,
so we must have $h=k+2$.$\square$

It is seen from this proof that
Theorem 4.2 will still hold true
even if property (i) is replaced
by a weaker property:

(i') There exists a vector $\bold d$ with positive real entries such that
$\bold d\chi_j<j\bold d$ componentwise.

\sl
Thus, we have shown that indecomposable $\Z_+$-representations of the
Verlinde algebra $V_k$ are in one-to-one correspondence with
Dynkin diagrams of finite type whose Coxeter number $h$ satisfies
the equation $k+2=h$.
\rm
\vskip .03in

4.3. Physical speculation.

We expect that the solution of the open problem in Chapter 3 is
related to the A-D-E classification of conformal field
theories with $\widehat{\frak sl}_2$-symmetry, due to Capelli,
Itzykson, and Zuber \cite{CIZ},\cite{Ka}. Their result is,
roughly speaking, that conformal field theories whose phase space
is a finite direct sum of spaces $L_i\otimes L_j^*$, where $L_i$ are
integrable highest weight
$\widehat{\frak sl}_2$-modules at level $k$, correspond
to simply laced Dynkin diagrams with Coxeter number $h=k+2$.
In particular, diagrams of type $A_{k+1}$ correspond
to the standard Wess-Zumino-Witten model, whose phase space
is $\oplus L_i\otimes L_i^*$. The fusion algebra of the WZW model
is $V_k$, and we know that the regular representation of
$V_k$, as a $\Z_+$-representation, also corresponds to the diagram
$A_{k+1}$. We believe that this correspondence should extend
to the diagrams of type $D$ and $E$, since the classifications
of conformal field theories and $\Z_+$-representations both correspond
to Dynkin diagrams, and the equation $k+2=h$ arises in both.
\vskip .03in

{\bf 5. Remarks.}

5.1. The notion of a $\Z_+$-algebra is closely related to
the notion of an integral table algebra introduced in \cite{B}.
An integral table algebra is a commutative $\Z_+$-algebra $A$ with unit
(the unit is in the basis)
with an automorphism of order at most $2$, denoted by *, such that:
(i) if $b$ is a basis element then $b^*$ is a basis element, (ii)
for any two basis elements $b^{(1)}$, $b^{(2)}$ the product $b^{(1)}b^{(2)}$
contains $1$ as a summand iff $b^{(1)}=(b^{(2)})^*$, and (iii)
there exists a homomorphism of algebras $d:A\to\C$
(the degree homomorphism) which takes
positive integer values on basis elements. For example, the character
ring of a compact group is an integral table algebra.

According to \cite{B}, a basis
 element $b$ of an integral table algebra $A$ is called
faithful if any basis element occurs in the decomposition of
$b^n$ for some $n$.

 In \cite{B}, integral table algebras with a faithful, self-dual
($b^*=b$) basis element $b$ of degree two are classified, and they turn out to
correspond to a subset of affine Dynkin diagrams, so that the matrix of
multiplication by $b$ is $2-A$, where $A$ is the Cartan matrix of the diagram.
This result is related Theorem 4.1 above. Indeed, every such integral table
algebra is automatically a $\Z_+$-representation of $(\Fun
SU(2))^{SU(2)}$, defined by $\chi_j\to P_j(b)$ (this follows from
Proposition 2.2). However, the condition for a $\Z_+$-representation
to be an integral table algebra with a faithful self-dual basis element of
degree 2 is not tautological, and thus not all diagrams are realized \cite{B}
($\tilde A_n$, $\tilde B_n$, $\tilde L_n$, $\widetilde{BL}_n$,
$\widetilde{CD}_n$, $\tilde{G}_{21}$ are not).

5.2. Algebras with nonnegative structure constants
and a $*$-automorphism (see above)
are known in the literature under the name table algebras
\cite{AB}, or commutative hypergroups.
Details on the theory of hypergroups can be found
in the review \cite{W}, and references therein. A-D-E classifications
similar to ours in Section 4.2 arise in the theory
of inclusion of factors (see \cite{GHJ}).

5.3. Theorem 4.1 can be regarded as a generalized McKay's
correspondence. It differs from the McKay's correspondence \cite{Mc}
by the appearance of non-simply laced diagrams. Other versions of
such a generalized correspondence are contained in \cite{Sl} and
\cite{EF}.

5.4. Examples of $\Z_+$-representations of the Verlinde algebra
are contained in \cite{Lu}, where the fusion algebras associated
to the non-abelian Fourier transform for dihedral Coxeter groups are
 constructed. For example, the algebra corresponding to the group
$I(5)$ described in \cite{Lu}, has the basis $1,p,q,r$ with relations
$qr=rq=p, q^2=q+1, r^2=r+1$, and has a structure of a
$\Z_+$-representation of the Verlinde algebra $V_3$ defined by
$\chi_2\to q$. This representation is equivalent to $L_2\oplus L_2$.

5.5. Happel, Preiser, and Ringel \cite{HPR} compute the
$\Z_+$-representations of $\Rep(SU(2))^{SU(2)}$ arising from
categories of representations of a finite subgroup in $SU(2)$ over a field
which is not algebraically closed. They show that it is possible
to obtain some non-simply laced Dynkin diagrams in this way,
but not all of them. The reason is that not every group of
automorphisms of a simply laced affine Dynkin diagram can be realized
by change of scalars.

{\bf Acknowledgements.} We are indebted to I.Frenkel, who explained to
us the categorical nature of $\Z_+$-algebras and representations. We are
grateful to W.Feit, I.Grojnowski, and G.Zuckerman for useful
discussions, information, and references.

\end
\enddocument